
%
\input harvmac
\def\coeff#1#2{\relax{\textstyle {#1 \over #2}}\displaystyle} 
\def\frac#1#2{{#1 \over #2}}
\def\shalf{\coeff{1}{2}}
\noblackbox
\lref\AFL{N. Andrei, K. Furuya, and J. Lowenstein, Rev. Mod. Phys. 55
(1983) 331; \hfill\break
A.M. Tsvelick and P.B. Wiegmann, Adv. Phys. 32 (1983) 453.}
\lref\PF{P. Fendley, Phys. Rev. Lett. 71 (1993) 2485,
cond-mat/9304031.}
\lref\FS{P. Fendley and H. Saleur, Nucl. Phys. B388 (1992) 609,
hep-th/9204094.}
\lref\FSii{P. Fendley and H. Saleur, ``Massless integrable quantum
field
theories and massless scattering in 1+1
dimensions'',  USC-93-022, hep-th/9310058.}
\lref\FSiii{P. Fendley and H. Saleur, ``Deriving Boundary $S$
Matrices'',
hep-th/\-9402045, to appear in Nucl. Phys. B.}
\lref\GZ{S. Ghoshal and A.B. Zamolodchikov, ``Boundary State and
Boundary $S$
Matrix in Two-Dimensional Integrable Field Theory'', RU-93-20,
hep-th/9306002.}
\lref\Card{J. Cardy, Nucl. Phys. B240 (1984) 514.}
\lref\ALii{I. Affleck and A. Ludwig, Phys. Rev. Lett. 67 (1991) 161.}
\lref\CK{C.G. Callan and I.R. Klebanov, Phys. Rev. Lett. 72 (1994) 1968,
hep-th/9311092.}
\lref\CKi{C.G. Callan, I. Klebanov, A.W.W. Ludwig and
 J. Maldacena, ``Exact solution of
a boundary conformal field theory'', PUPT-1450, IASSNS-HEP-94/15,
 hep-th/9402113.}
\lref\SK{E.K. Sklyanin, J.Phys. A21 (1988) 2375.}
\lref\Zamo{Al.B. Zamolodchikov, Phys. Lett. B253 (1991) 391.}
\lref\PT{J. Polchinski and L. Thorlacius, ``Free fermion representation
of a boundary conformal field theory'', NSF-ITP-94-29, hep-th/9404008.}
\lref\ZZI{A.B. Zamolodchikov and Al.B.
Zamolodchikov, Nucl. Phys.  B379 (1992) 602.}
\lref\ZI{Al.B. Zamolodchikov, Nucl. Phys. B358 (1991) 524.}
\lref\FT{L.D. Faddeev and L.A. Takhtajan, Phys. Lett.  85A (1981) 375.}
\lref\YY{C.N. Yang and C.P. Yang, J. Math.  Phys. 10 (1969) 1115.}
\lref\Ztba{Al.B. Zamolodchikov, Nucl. Phys. B342 (1991) 695.}
\lref\JW{R. Jackiw and G. Woo, Phys. Rev. D12 (1975) 1643.}
\lref\Kor{V.E. Korepin, Th. Math. Phys. 34 (1978) 1.}
\lref\Ablo{M. Ablowitz and H. Segur, {\it Solitons and
the inverse scattering transform}, Philadelphia: SIAM, 1981}
\lref\ghosi{S. Ghoshal, ``Bound State Boundary $S$ Matrix of the
Sine-Gordon
Model'',  RU-93-51, hep-th/9310188.}
\lref\Guin{F. Guinea, Phys. Rev. B32 (1985) 7518.}
\lref\Schmid{A. Schmid, Phys. Rev. Lett. 51 (1983) 1506; \hfill\break
F. Guinea,
V. Hakim and A. Muramatsu, Phys. Rev. Lett. 54 (1985) 263.}
\lref\Melzer{E. Melzer, ``Massive-Conformal Dictionary'',
TAUP\--2109\--93,
 hep\--th/\-9311\-058.}
\lref\BCNA{H.W. Blote, J.L. Cardy and M.P. Nightingale, Phys. Rev.
Lett. 56
(1986) 742; I. Affleck, Phys. Rev. Lett. 56 (1980) 746.}
\lref\TS{M. Takahashi and M. Suzuki, Prog. Th. Phys. 48 (1972) 2187.}
\lref\KM{T.R. Klassen and E. Melzer, Nucl. Phys. B338 (1990) 485.}
\lref\Kore{V.E. Korepin, Th. Math. Phys. 41 (1979) 953.}
\lref\FST{L. Faddeev,
E. Sklyanin and L. Takhtajan, Th. Math. Phys. 40 (1979) 688.}
\lref\KF{C.L. Kane and M.P.A. Fisher, Phys. Rev. B46 (1992) 15233.}
\lref\FLS{P. Fendley, A. Ludwig and H. Saleur, in preparation.}
\lref\QHE{X.G. Wen, Phys. Rev. B41 (1990) 12838;
K. Moon, H. Yi,  C.L. Kane,
S.M. Girvin, and M.P.A. Fisher, Phys. Rev. Lett. 71 (1993) 4381.}
\lref\Mccoy{S. Dasmahapatra, R. Kedem, T.R.
Klassen, B.M. McCoy and E. Melzer, ``Quasiparticles, Conformal Field
Theory,
and q-series'',  hep-th/9303013.}
\lref\SY{R. Sasaki and
I. Yamanaka, Adv. Stud. in Pure Math. 16 (1988) 271.}
\lref\EA{S. Eggert and I. Affleck, Phys. Rev. B46 (1993) 10866.}

\def\t{\theta}
\def\l{\lambda}

\def\<{\langle}
\def\>{\rangle}
\Title{\vbox{\baselineskip12pt
\hbox{USC-94-10}\hbox{hep-th/yymmxxx}}}
{\vbox{\centerline{Exact solution of a massless scalar field}
\vskip4pt\centerline{with a relevant boundary interaction}}}

\centerline{P. Fendley, H. Saleur$^*$ and N.P. Warner}
\bigskip
\centerline{Department of Physics, University of Southern California}
\centerline{Los Angeles CA 90089-0484}
\vskip.3in
We solve exactly the ``boundary sine-Gordon'' system of a massless
scalar field $\phi$ with a $\cos\beta\phi/2$ potential at a boundary.
This model has appeared in several contexts, including tunneling between
quantum-Hall edge states and in dissipative quantum mechanics.
For $\beta^2 <8\pi$, this system exhibits a boundary
renormalization-group flow from Neumann to Dirichlet
boundary conditions.
By taking the massless limit of the sine-Gordon model
with boundary potential, we find the exact $S$ matrix for
particles scattering off the boundary. Using the thermodynamic Bethe
ansatz, we calculate the boundary entropy along the entire flow.
We show  how these particles correspond to  wave packets in the
classical Klein-Gordon equation, thus giving a more precise explanation
of scattering in a massless theory.

\bigskip
\bigskip\bigskip
\noindent $^*$ Packard Fellow
\Date{6/94}

\newsec{Introduction}

Many fundamental models of statistical mechanics can be expressed as
$1+1$-dimensional field theories with a boundary. A number
of these field theories
can be exactly solved by applying the powerful techniques of
integrability \AFL.
Such models include the Kondo problem, a number of
models of dissipative quantum mechanics, and the model we treat
here: a free boson with a periodic potential on the boundary.

We consider the ``boundary sine-Gordon'' problem of a massless
scalar field $\phi(\sigma)$ on the segment
$0<\sigma<L$.
The Lagrangian is
\eqn\lagr{{\cal L}={1\over 2}\int_0^L
d\sigma\left(\partial_\mu\phi\right)^2 +
g\cos\left[{\beta\over 2}\phi(0) \right].}
This model arises in a number of contexts because one can
often map interacting fermions (the Luttinger liquid or
massless Thirring model) onto a free boson. Coupling the fermions
to a boundary potential, for example, gives this boundary interaction
for the bosons \KF.
This model has also appeared in dissipative quantum mechanics,
where it describes a particle moving in a periodic potential
with a frictional force \Schmid.

The model in the bulk (a free massless scalar field) is, of course,
a conformal field theory.
For $g$=0 and $g\to\infty$, the combination
of left
and right conformal symmetries is preserved in the presence of a
boundary
\Card. However for $\beta^2<8\pi$,
the interaction breaks the scale invariance,
with the parameter $g$ providing a mass scale.
At $g$=0, the boundary dimension of the operator $\cos[\beta\phi/2]$ is
$x={\beta^2\over 8\pi}$.
For 0$<x<$1 a perturbation by this operator is relevant and induces a
renormalization group flow between two different conformal boundary
conditions.
Our purpose is to study this flow explicitly.

To understand the boundary condition induced by the interaction, we
examine the energy-momentum tensor
$T_{\mu\nu}=\half\del_\mu\phi\del_\nu\phi$,
which satisfies the current conservation $\del_{\mu} T_{\mu\nu}=0$. The
Hamiltonian is defined as $H=\int_0^L T_{tt}d\sigma
- g\cos[\beta\phi(0)/2]$.
The interaction adds a boundary contribution to $H$, so that the
boundary
itself can hold energy.
Without the boundary interaction, we would require that the component
$T_{\sigma t}$ vanish at
$\sigma=0$ and $\sigma= L$
to ensure that no energy crosses the boundaries (i.e. $dH/dt=0$).
 Thus in general,
energy conservation gives a dynamical boundary condition, requiring
that
$T_{\sigma t}=\del_t g\cos[\beta\phi/2]$ at $\sigma=0$, or
\eqn\bdry{{d\phi(0)\over d\sigma}={g\beta}
\sin\left[{\beta\over 2}\phi(0)\right].}
This is a classical equation, but the same relation holds in the quantum
field theory after a renormalization of parameters \GZ.
Thus the boundary interaction causes a flow between Neumann boundary
conditions ($\del\phi(0)=0$) in the UV and Dirichlet
($\phi(0)=4n\pi/\beta$, $n$
integer)
in the IR. These two boundary conditions are dual to each other.

To solve this model, we shall make use of the powerful constraints of
integrability. There are an infinite number of conserved kinematic
quantities
for the free boson. A generic boundary interaction will destroy them.
However,
a boundary interaction of the form \lagr\ (or equivalently, a boundary
condition of the form \bdry) still preserves an infinite number of
appropriately-modified conserved quantites. This can be seen explicitly
in
perturbation theory and in the classical limit, by taking the massless
limit of
the sine-Gordon model with a boundary \GZ.

\nref\ZZ{A.B. Zamolodchikov and Al.B. Zamolodchikov, Ann. Phys. 120
(1979) 253.}
\nref\ZAdv{A.B. Zamolodchikov, Adv. Stud. Pure Math. 19 (1989) 1. }
\nref\Muss{G. Mussardo, Phys. Rep. 218 (1992) 215.}
A particularly useful approach to integrable theories with and without
boundaries is to find the exact $S$ matrix of the physical
quasiparticles of
the theory \refs{\ZZ-\Muss}. From the $S$ matrix one can calculate a
number of
exact thermodynamic quantities such as the free energy, specific heat,
susceptibility
and boundary entropy. This is done by using the thermodynamic Bethe
ansatz
\refs{\YY,\Ztba}. We describe this technique in the presence of a
boundary
in sect.\ 4.

To understand the states of our theory, we look at the system far away
from the
boundary. Here the ``bulk'' theory is just that of a free massless
scalar field.
Ordinarily, with a massless bulk theory, one describes the excitations
in terms of plane waves (perhaps around some non-trivial topological
background). However, this is not a convenient basis to work with once
one has
included the boundary interaction. The reason for this is simple. The
plane
waves are no longer eigenstates of the conserved charges, so a plane
wave
scattering off the boundary could result in a possibly-horrible
combination of
other plane waves.

A basis where the states have well-defined eigenvalues of the conserved
charges
is given by the massless limit of sine-Gordon solitons \FSii. In other
words,
we consider therefore the problem
given by the Lagrangian \lagr\ as the $G\to 0$ limit of a problem
which is massive in the bulk,
with perturbation $\delta {\cal L}=-G\int_0^L d\sigma \cos\beta\phi$.
With this particular combination of bulk and boundary perturbations
(notice the
$\beta$ in the bulk and the $\beta/2$ on the boundary) the model is
integrable
\GZ. In sect.\ 2, we will
make this change-of-basis explicit by showing that the massless
solitons obey the classical equation \bdry, and that they retain their
form after scattering.

When one changes basis from the free plane waves to the massless
sine-Gordon
solitons, the scattering of the bulk particles is no longer trivial or
even
diagonal. We describe such processes by
massless scattering \refs{\ZI,\ZZI,\FSii}.
The integrability of the model is crucial to such a description. In a
basis with
well-defined charges, all collisions must be completely elastic, so
that
momenta are conserved individually and particle production is
impossible.
Even with this crucial simplification, it is still not immediately
obvious what
it means to scatter two
particles traveling in the same direction at the speed of light.
There are several ways to define a bulk $S$ matrix in the integrable
models we
are studying. One is as the phase which results when commuting two
particle-creation operators (the Zamolodchikov-Faddeev algebra). This
phase is
equivalently realized as a matching condition on a two-particle
wavefunction
$\psi(\sigma_1,\sigma_2)$ between the regions $\sigma_1\ll \sigma_2$
and
$\sigma_1\gg \sigma_2$. One can explicitly compute this phase in a
lattice
Bethe ansatz by forming a two-particle state and bringing one of the
two
particles around the periodic world, i.e.\ sending $\sigma_1 \to
\sigma_1 +L$,
where space is a circle of circumference $L$ \FT. In other words, one
studies
the monodromy properties of massless particles \ref\RS{N.Yu  Reshetikhin and H.
Saleur, ``Lattice regularization of massive and massless integrable field
theories'', USC-93-020, hep-th/9309135.}.

We emphasize that by ``massless'' limit, we mean that the
quasiparticles are
massless (they have the dispersion relation $E=\pm P$). The theory is
not
scale-invariant because of the boundary interaction. With this scale
(we call
it $T_B$), boundary scattering can depend on the momentum of the
incident
particle, because there exists a dimensionless parameter $P/T_B$.
Similarly,
thermodynamic quantities like the free energy
depend on the ratio $T/T_B$.

The outline of this paper is as follows. In sect.\ 2, we discuss the
classical
limit of the problem in order to give a more precise understanding of
massless
scattering. While it is conceptually important, the details are not
required
for what follows. In sect.\ 3, we give the exact quantum $S$ matrix by
taking
the massless limit of the boundary $S$ matrix for the sine-Gordon model
\GZ. We
show that it has all of the desired properties, and that it agrees with
a
previously-derived result at $\beta^2=4\pi$ \Guin. In sect.\ 4, we use
the $S$
matrix to derive the boundary free energy, thus showing how the
boundary
entropy flows between the Neumann and Dirichlet fixed points. In sect.\
5 we
present some thoughts on future directions.


\newsec{Understanding massless scattering via the classical limit}

Studying the scattering of massless particles has given a wealth of
useful
information generally unavailable by other means; for example, as we
discuss in
sect.\ 4, it enables the calculation of the exact free
energy ($c$-function) flowing between
two fixed points (for a review, see \FSii).
One subtlety is that in such a description
there can be non-trivial bulk scattering even if
a massless theory is linear and obeys
superposition, and thus
appears to have trivial monodromy. For example, in the Kondo problem,
the bulk
model consists of free massless electrons; in this paper we discuss a
free
massless scalar field.
The reason for the non-trivial scattering, as indicated in the
introduction, is that when coupling to a boundary, it is
much more useful to make a (non-linear) change of basis of particles.
The
advantage is that boundary scattering is much easier to describe; the
added
complication is that the scattering in the bulk is no longer trivial.
Thus in the multi-channel Kondo problem, the appropriate quasiparticles
are
kinks in a potential with mutiple degenerate wells \PF, whereas in this
paper
they are kinks (and breathers) in a potential with an infinite number of
wells.
This change-of-basis is often not very explicit, since it can be
deduced
indirectly from the Bethe ansatz solution, or inferred from demanding
the
appropriate symmetry structure.
Our purpose in this section is to elucidate
massless scattering and hopefully give some
physical insight by describing the classical limit
of an important example.

We consider here
a classical scalar field $\phi(\sigma,t)$ satisfying the
Klein-Gordon equation:
\eqn\KGeqn{\partial_t^2 \phi ~-~ \partial_\sigma^2 \phi ~=~ 0 \ .}
A dimensional parameter $g$ is introduced by requiring that theory live
on the
half-line $[0,\infty)$, with boundary conditions:
\eqn\intbcs{\del_\sigma \phi \big|_{\sigma=0} ~=~ g \sin\big(
\shalf(\phi -
\phi_0)
\big) \big|_{\sigma=0} \ ,}
for some constants $g$ and $\phi_0$.  (Classically, one can scale out
the
parameter $\beta$ in \bdry.)

\subsec{The massless limit of sine-Gordon}

In many massless scattering
problems an intrinsic bulk massive
perturbation has been implicitly or
explicitly introduced, and then set to zero.
Moreover, the perturbation is always
chosen so that the theory remains integrable off the critical point.
The fact that one is
considering a massless limit of a massive integrable model leads one
to a preferred basis of `wave packets' or `wavelets' for the massless
theory.
This enables one to give a particle interpretation to the massless
theory, and defines the concept of massless scattering.
A given massless theory can, of course, be the limit of several massive
integrable theories (e.g.\ the Ising model with
a magnetic or energy perturbation), and as a result there
can be several different massless scattering matrices, leading
to distinct descriptions of the massless theory (see, for
example, \Mccoy). A bulk mass can be introduced to our example by
considering
\KGeqn\ to be the
$m \to 0$ limit of the sine-Gordon equation:
\eqn\SGeqn{\partial_t^2 \phi ~-~ \partial_\sigma^2 \phi ~=~ -m^2
\sin(\phi) \ .}
This massive model was shown to be integrable in the presence of the
boundary
condition \intbcs\ in \GZ.

For either of these partial differential equations we want to consider
wave
packets of finite energy (which means that $\phi$ must
approach a constant value suitably fast as $\sigma \to \pm \infty$).
For any non-linear, integrable, partial differential equation, there
are infinitely many conserved quantities, and in the limit in which
the mass vanishes, these conserved quantities become conserved
quantities
for the massless theory.  For the sine-Gordon theory, the massless
limit
of these conserved quantities are special polynomials in the
partial derivatives of $\phi$.  The most general characterization of
the
preferred wave packets in the massless theory are those wave packets
that
are eigenfunctions of the massless limit of these conserved quantities.
An equivalent characterization of these massless wave packets is to
take the non-dispersive wave packets
of the massive theories and take an appropriate
limit in which their energy remains finite when $m\to 0$.

There are two types of finite-energy solutions of the classical
sine-Gordon
equation \SGeqn: solitons, which are time-independent and topologically
non-trivial, and breathers, which are time-dependent and topologically
trivial.
Intuitively, a breather can be thought of as a bound state of a kink
and an
antikink oscillating in and out (i.e.\ breathing).
In this section, we will discuss only the solitons; the analysis for
the
breathers follows analogously.

A major triumph of the theory of non-linear partial differential
equations was
the construction of explicit solutions of \SGeqn\ for any number of
moving
solitons (see, for example, \Ablo).
The solitons' energies and momenta
are conveniently expressed
in terms of rapidities $\alpha_j$, defined by
$E_j=m\cosh\alpha_j$ and $P_j=m\sinh\alpha_j$. The velocity of each is
thus
given by $\tanh \alpha_j$ (positive for a right-moving soliton).
We have set the speed of ``light'' to be 1.

To find the scattering of two solitons in the massless limit,
consider a two-soliton solution of \SGeqn\ on
$(-\infty,\infty)$.  This solution is usually expressed as:
\eqn\sgsoln{\phi(\sigma,t) ~=~ 4 ~ arg (\tau) ~\equiv~ 4~ \arctan\bigg(
{{\cal I}m(\tau)  \over {\cal R}e (\tau) }  \bigg) \ ,}
where the $\tau$-function solution is given by:
\eqn\twosols{\eqalign{\tau ~=~ 1 ~-~ & \epsilon_1 \epsilon_2
\left( \tanh{\alpha_1-\alpha_2 \over 2}\right)^2 ~e^{ -
E_1 (\sigma-a) - E_2 (\sigma-b) +P_1 t +P_2 t}  \cr
{}~+~ & i \big\{ ~ \epsilon_1 e^{- E_1 (\sigma-a)+P_1 t}
{}~+~ \epsilon_2 e^{- E_2 (\sigma-b)+P_2 t} ~\big\}
 \ , \cr}}
The constants
$a$ and $b$ represent the initial positions of the two solitons, and
$\epsilon_j = +1$ if the $j^{\rm th}$ soliton is a kink, while
$\epsilon_j = -1$ if it is an anti-kink.

For a wavepacket to have finite energy in the massless limit $m\to 0$,
the rapidity $|\alpha|$ must go to infinity. We thus define
$\alpha\equiv\Lambda+\t$, and let $\Lambda\to\infty$ such that the
parameter
$\mu\equiv \shalf me^{\Lambda}$ remains finite. The energy and momentum
of a
right-moving ``massless'' soliton then reads
$E=P=\mu e^\t$. For a left mover, $\alpha\equiv -\Lambda+\t$, and its
energy
and momentum read $E=-P=\mu e^{-\t}.$

Suppose that both of these solitons are right-moving. Then the massless
limit yields:
\eqn\infrapid{\tau ~=~ 1 ~-~ \epsilon_1 \epsilon_2
e^{-\Delta}e^{ -E_1(\eta - a) -E_2(\eta  - b) }
{}~+~  i \big\{ ~ \epsilon_1 e^{- {E_1 (\eta - a)} }
{}~+~ \epsilon_2 e^{- { E_2 (\eta - b)}} ~\big\}}
where $\eta = (\sigma-t)$ and
$$\Delta ~\equiv~  - ~\log\big[(\tanh(\theta_1 - \theta_2))^2\big].$$
This is manifestly a solution of the
Klein-Gordon equation, and by construction is an eigenfunction of all
of the conserved quantities of the massless limit of the sine-Gordon
equation. (It is shown in \FSii\
that it remains an eigenfunction after
quantization --- the expectation values $\<E^{2k-1}\>$ are
precisely the quantum conserved charges derived in \SY.) It is almost a
superposition of two single-soliton
wave packets.  Observe that:
\eqn\superpos{\eqalign{ arg \big[ 1 ~-~ \epsilon_1 &\epsilon_2
 ~e^{ - E_1(\eta - a) - E_2(\eta  - b) }
{}~+~  i \big\{ ~ \epsilon_1 e^{- E_1 (\eta - a) }
{}~+~ \epsilon_2  e^{-  E_2 (\eta - b)} ~\big\} \big] \cr
{}~=&~ arg \big[ 1 + i \epsilon_1 e^{- E_1 (\eta - a) }  \big]
{}~+~ arg \big[ 1 + i \epsilon_2 e^{- E_2 (\eta - b) }  \big]
\ .}}
The factor $\Delta$ thus measures
the extent to which the two-soliton solution is not a superposition
of one-soliton solutions.

More precisely, consider the limit $a \to \infty, \eta \to \infty$
so that $E_1(\eta - a)$ is finite.  This corresponds to moving the
first
kink off to $\sigma=+\infty$ and following it.  The $\tau$ function
collapses
to the one-kink form $\tau = 1 + i\epsilon_1 e^{- E_1 (\eta - a)} $.
Moving this soliton through the second one corresponds to taking it
to $\sigma = -\infty$, or taking the limit $a \to - \infty, \eta \to
-\infty$
(with  $E_1(\eta - a)$ finite).  Discarding an overall multiplicative
factor (which is irrelevant in the computation of $\phi = 4
\arg(\tau)$),
we see that in this limit, $\tau = 1 + i\epsilon_1 e^{- E_1 (\eta -
a) -\Delta}$.
Thus these preferred Klein-Gordon wave packets exhibit
non-trivial monodromy.  The foregoing time delay $\Delta$ is precisely
the classical form of a massless scattering matrix. One can then
semi-classically quantize this \refs{\JW,\Kor}; the analysis is
identical in
 massive and massless cases, because the difference $\t_1-\t_2$ is the
same in
both cases. We will bypass this step because the exact quantum answer
is
already known \ZZ.

One obtains the same $\Delta$ for two left-moving
solitons.   For a left-moving and a right-moving soliton colliding
one easily sees that the massless limit of $\big(
\tanh\shalf (\alpha_1-\alpha_2)\big)^2$ is unity.  The solution
collapses
to the superposition of a left-moving wave packet and a right-moving
 wave packet exactly as in \superpos, with no time delay.  This
is the classical manifestation of the fact that the left-right quantum
scattering matrix $S_{LR}$ elements are at most rapidity-independent
phase
shifts.

\subsec{Klein-Gordon with a non-linear integrable boundary condition}

Consider now the Klein-Gordon equation on $[0,\infty)$ with the
boundary
condition \intbcs.  The non-linear boundary conditions destroy the
naive conserved quantities of the Klein-Gordon equation.  However, by
reversing the argument in the
appendix of \GZ\ one finds that the system
will still have higher-spin conserved quantities, namely those of the
massive sine-Gordon system.  Thus the bulk theory is
Klein-Gordon, but in this instance there are a preferred set of wave
packets
that scatter from the boundary without dispersion.  These wave packets
are once again the massless limit of the sine-Gordon solitons.

A more direct way of seeing the integrability of the Klein-Gordon (and
indeed, sine-Gordon) equation with boundary conditions \intbcs\ is
provided by the analysis of \ref\HSSSNW{H.~Saleur, S.~Skorik and
N.P.~Warner, in preparation.}.
The idea is to show that the method of images can
be used on $(-\infty,\infty)$ even in the non-linear system, so as to
replicate the boundary conditions \intbcs\ on $[0,\infty)$.
The scattering of a kink, or anti-kink, from the boundary can be
described  by a three-soliton solution on $(-\infty,\infty)$. These
three solitons consist of the incoming soliton, its mirror image
with equal but opposite velocity, and
a stationary soliton at the origin (to adjust the boundary conditions).
If one takes the infinite rapidity limit of this three-soliton solution
then
the stationary soliton simply collapses to an overall shift of $\phi$
by a constant, while the mirror images (since they are moving in
opposite
directions) reduce to a superposition of two wave packets.  One thus
obtains:
\eqn\bndryscatt{\phi ~=~ \phi_0 ~+~ 4~ arg \big[ 1 + i \epsilon_1 e^{-
{E (\xi - a)} }  \big] ~+~ 4~ arg \big[ 1 + i \epsilon_2
e^{-
{E (\eta - b)} }  \big] \ ,}
where $\xi = \sigma+t$, $\eta = \sigma - t$.  By direct computation
one finds that this solution satifies \intbcs\ with:
\eqn\twosolbc{ e^{E(a+b)}
{}~=~ - \epsilon_1 \epsilon_2 ~ {(2E+g) \over (2E-g)} \ .}
The constant $\Delta_B\equiv -E(a+b)$ represents
the delay of the reflected pulse.
If one defines the classical boundary scale $\t_B$ via $g = 2\mu
e^{-\t_B}$,
then this delay
may be written as
\eqn\refdelay{\Delta_B ~=~ \log (- \epsilon_1 \epsilon_2
{}~ \tanh \shalf (\theta - \t_B)) \ .}
Note that the sign  $ \epsilon_2$ is to be chosen so as to make
the argument of the logarithm real.  This determines whether the
reflection of a kink will be a kink or an anti-kink. Thus we see that
$\t_B$ is
the scale at which behavior crosses over from the region of the Neumann
critical point (where the classical boundary scattering is completely
off-diagonal) to the Dirichlet boundary critical point (where classical
boundary
scattering is diagonal).
This result is completely consistent with the infinite-rapidity
limit of the
delay computed in \HSSSNW.

Before concluding this section,
we wish to observe that the foregoing is not simply
a property of some solitonic solutions, but is a general feature of the
sine-Gordon
equation and its massless limit.  The sine-Gordon equation (and
any other integrable equation) admits B\"acklund transformations.
In particular, consider
\eqn\Backlund{ \partial_\eta (\phi ~+~ \psi) ~=~ mD~
\sin \Big({\phi - \psi \over 2} \Big) \ ;  \qquad
\partial_\xi (\phi ~-~ \psi) ~=~ {m \over D}~
\sin \Big({\phi + \psi \over 2} \Big) \ ,}
where $D$ is an arbitrary parameter.  The point is that $\phi$
satisfies
the sine-Gordon equation if and only if $\psi$ does so as well.
Taking $m\to 0$ while keeping $mD$ finite, the
B\"acklund transformation becomes:
\eqn\scaledBacklund{ \partial_\eta (\phi ~+~ \psi) ~=~ m D~
\sin \Big({\phi - \psi \over 2} \Big) \ ;  \qquad
\partial_\xi (\phi ~-~ \psi) ~=~ 0 \ .}
One can easily check that this transformation implies that $\phi$
satisfies
the Klein-Gordon equation if and only if $\psi$ does so as well.
However,
suppose that $\psi$ is a solution to Klein-Gordon on $[0,\infty)$
satisfying
Dirichlet boundary conditions $\phi = \phi_0$.  It follows
immediately from \scaledBacklund\ that $\psi$ is a solution to
Klein-Gordon satisfying \intbcs\ with $g = - mD$.
For any left moving wave-packet, $f(\xi)$, the Dirichlet scattering
solution $\phi$ is given by $\phi = \phi_0 + f(\xi) - f(-\eta)$.
The scattering solution, $\psi$, for the same incoming wave-packet
reflecting from the boundary conditions \intbcs\ can then, in
principle,
be found by by solving \scaledBacklund.  That is, if one writes  $\psi
= f(\xi) + r(\eta)$ then $r$ satisfies $r'(\eta) = f'(\eta) + g \sin
(({r(\eta) + f(-\eta) - \phi_0}) / 2)$.

\newsec{The quantum $S$ matrix}

We find the quantum $S$ matrix in the same manner as the classical, by
taking
the massless limit ($G\to 0$) of the sine-Gordon $S$ matrix \GZ. The
Lagrangian
for the massive theory is
\eqn\SGlagr{{\cal L}={1\over 2}\int_0^L d\sigma\left[
\left(\partial_\mu\phi\right)^2+ G \cos\beta\phi\right]
+g\cos\left[{\beta\over 2}\phi(0)\right].}
The physics of the model depends crucially on the parameter $\beta$,
which can
be scaled out when studying the classical equations of motion.
It is convenient to define the parameter
$$\lambda\equiv {8\pi\over\beta^2}-1.$$
In the
classical
limit $\beta \to 0$, the results of this section of course reduce to
those of
the previous section. We should note that a massless $S$ matrix can
often be
derived directly, without taking a limit of a massive theory
\refs{\ZI,\ZZI,\FSii,\PF}. This is done by imposing the constraints of
the
Yang-Baxter equation, crossing and unitarity.

\subsec{The massless bulk $S$ matrix}

The bulk
structure of the quantum sine-Gordon theory is very well known
\refs{\ZZ,\FST,\Kore}. At any
value of $\beta$, the spectrum includes a soliton $S$ and an
antisoliton $A$
of mass $m$,
with $m\propto G^{(\l+1)/2\l}$. These are the quantized particles
corresponding
to the classical solitons discussed in the previous section. Moreover,
we have
bound states (breathers) indexed by $n$ a positive integer less than
$\l$, with
 mass $m_n=2m\sin n\pi
/2\lambda$. As for the classical theory, we obtain the massless limit
by sending the rapidities of these particles
to $\pm \infty$ and at the same time scaling appropriately the mass
$m\to 0$ so that the energy of each particle remains finite.
In this limit we obtain a set of left-moving
and right-moving particles. The right movers have dispersion relation
$E=P$ and
can be parametrized as
\eqn\disp{E_{S,A}=\mu e^\t \qquad
E_n=2\mu\sin\left({n\pi\over 2\lambda}\right)e^\t,}
where $\mu$ and $\theta$ are now renormalized mass and rapidity. The
left
movers have $E_{S,A}=-P=\mu e^{-\t}$ and likewise for the breathers.

The left-left and right-right two-particle $S$ matrices are given by
the same
formula as
in the massive case. The matrices $S_{LL}$ and $S_{RR}$ depend on the
difference
of renormalized rapidities because the only relativistic invariant that
can be formed is the ratio of momenta. (There are no dimensional
parameters in
the bulk problem). The $S$ matrix is purely elastic; individual momenta
do
not change in a collision.
Although the classical $S$ matrix is diagonal, the quantum
one is not: the initial state $|S(\t_1)A(\t_2)\>$ can scatter to
$|A(\t_1)S(\t_2)\>$ because $A$ and $S$ have the same mass.
For the soliton and antisoliton one has
the three usual amplitudes \ZZ
\eqn\sll{\eqalign{a(\t)&=\sin\lambda(\pi+i\t)Z(\t)\cr
b(\t)&=-\sin\lambda i\t Z(\t)\cr
c(\t)&=\sin\lambda\pi Z(\t).\cr}}
where the element $a(\t_1-\t_2)$ describes the process
$|S(\t_1)S(\t_2)\>
\rightarrow |S(\t_1)S(\t_2)\>$, $b$ describes $SA\to SA$, $c$ describes
the
non-diagonal process $SA\to AS$, and there is a symmetry under
interchange of
soliton to antisoliton (corresponding to $\phi\to -\phi$).  The
function
$Z(\t)$ is a well-known normalization factor, which can be written as
$$Z(\t)={1\over \sin\l(\pi +i\t)}\exp\left(i
\int_{-\infty}^{\infty} {dy\over 2y} \sin {2\t y\l \over\pi} {\sinh
(\l-1)y
 \over \sinh y\cosh \l y} \right).$$
Notice that when $\l$ is an integer, the scattering is diagonal and
that $a=\pm
b$.

The left-right scattering comes from the $\t\to\infty$ limit of \sll;
it is
diagonal and rapidity-independent, but this constant phase is different
for
$SS\to SS$ and $SA\to SA$. A non-constant $S_{LR}$ induces a flow
between bulk
critical points \refs{\ZI,\ZZI,\FSii}; we will not discuss this here.

The breather-soliton and breather-breather $S$ matrices are well known
\ZZ. The
massless limit is taken in the same way, so $S_{LL}$ and $S_{RR}$  are
the same
as the massive $S$ matrix, while $S_{LR}$ is a constant.

\subsec{The boundary $S$ matrix for solitons}

To study the effect of the dynamical boundary condition we now consider
the
scattering of these massless particles off the boundary. Since the
boundary
introduces a scale $T_B$ (the ``boundary temperature'') to the problem,
the
boundary $S$ matrix element for a particle with momentum $P$ depends on
the
ratio $P/T_B$. Defining the boundary scale $\t_B$ via
$T_B\equiv \mu e^{-\t_B}$, we see that the $S$ matrix element for a
left mover
to scatter off the boundary (and thus become a right mover) depends on
$\t-\t_B$.

We can obtain
the $S$ matrix from the results of \GZ.
This $S$ matrix has two free parameters, $\eta$ and $\Theta$. Setting
$\phi_0=0$ in the boundary potential
as we have done corresponds to setting $\eta=0$.
The remaining parameter $\Theta$ is a
function of the boundary scale $g$.
The $S$ matrix then depends on a product of functions of
$\alpha,\lambda\alpha+\Theta$, and $\l\alpha-\Theta$,
where $\alpha$ here is the
unrenormalized
rapidity (which is called $\theta$ in \GZ).
To get the scattering of left movers off of the boundary, we simply
take the
limit
of very large and negative $\alpha$ and appropriately
scale the
parameter $\Theta$, so that $\alpha-(\Theta/\l)\equiv\t-\t_B$ remains
finite.
The effect of this
scaling is to obtain left-moving particles of energy comparable to the
boundary
term energy.

There are two different boundary amplitudes:
$P(\t-\t_B)$ for the $S \rightarrow S$ and $A\to A$ processes, and
$Q(\t-\t_B)$ for $S \rightarrow A$ and $A\to S$.
The symmetry
$\phi \to -\phi$ ensures that the amplitudes are the same when soliton
and
antisoliton are interchanged. This $S$ matrix is not diagonal; notice
that the
boundary perturbation breaks the $U(1)$ soliton-number symmetry, thus
allowing
soliton to scatter into an antisoliton at the boundary. As
$g\to\infty$, this
symmetry is restored, so at the Dirichlet fixed point, the element $Q$
must
vanish.
Taking the limit of the result of \GZ\ as described, one finds
\eqn\bdrs{\eqalign{P(\t)&={e^{\lambda\t/2}}R(\t)\cr
Q(\t)&=i{e^{-\lambda\t/2}}R(\t)}}
where the function $R$ reads
$$\eqalign{R(\t)&={e^{i\gamma}
\over 2\cosh({\lambda \t\over 2}-i{\pi\over 4}) }
\prod_{l=0}^\infty{Y_l(\t)\over Y_l(-\t)}\cr
Y_l(\t)&={\Gamma\left({3\over 4}+l\lambda
-{\lambda i\t\over 2\pi}\right)\Gamma\left({1\over 4}+(l+1)\lambda
-{\lambda i\t\over 2\pi}\right)\over \Gamma\left({1\over 4}+(l+\half)
\lambda -{\lambda i\t\over 2\pi}\right)\Gamma\left({3\over 4}+(l+\half)
\lambda -{\lambda i\t\over 2\pi}\right)}.\cr}$$
A useful integral representation of $R$ is given by
\eqn\norm{R(\t)={e^{i\gamma}
\over 2\cosh({\lambda \t\over 2}-i{\pi\over 4}) }
\exp i\int_{-\infty}^{\infty} {dy\over 2y} \sin{2\l\t y\over \pi}
{\sinh(\l-1)y \over \sinh 2y\cosh \l y }}
The constant phase $\gamma$ depends on $\lambda$ and turns out to be
necessary
to satisfy the cross-unitarity relation \GZ. For right movers
scattering off a
boundary, the answer is similar but differs in some constant phases and
signs.

We can check this answer in several limits. As $\t_B\to \infty$, we
obtain the
Neumann fixed point. Since $P(\t-\t_B)\to 0$ in this limit,
the boundary scattering is completely off-diagonal, so that
the $U(1)$ soliton-number symmetry is maximally violated.
As $\t_B\to -\infty$,
we reach the Dirichlet fixed point, where the scattering is diagonal.
The results
at both limits follow from the analysis of \KF.

We can also check the answer at several values of $\lambda$. The case
$\lambda=1$, where the perturbing operator has dimension $\half$,
has already been solved \Guin. It was shown
explicitly in the corresponding lattice model that the bulk theory
is equivalent to two critical Ising models, and that the boundary
interaction corresponds to a boundary magnetic field in one
of the two Ising models. We have the same result here.
The bulk theory corresponds to a free massless Dirac fermion, where the
soliton
corresponds to the fermion and the antisoliton the antifermion.
This is easily seen in \sll: the $S$ matrix elements are $a=b=1$,
$c=0$. We can
decompose this into the massless Majorana fermions of two critical
Ising models
by looking at the linear combinations $(|S\>+|A\>)$ and $(|S\>-|A\>)$.
The
corresponding boundary $S$ matrix elements are then $P+Q$ and $P-Q$.
For
$\lambda=1$, we then have
$$\eqalign{P+Q&=1\cr
P-Q&=\tanh\left({\t\over 2}-{i\pi\over 4}\right)\cr}$$
The $S$ matrix for an Ising model in a boundary magnetic field indeed
is
$\tanh\left({\t\over 2}-{i\pi\over 4}\right)$, as can be seen by explicit
calculation
or by taking the limit of the massive $S$ matrix in \GZ.

Another interesting value is $\l=0$, where the dimension of the
perturbing
operator is $1$, so that the operators $\exp(i4\pi\phi),\
\exp(-i4\pi\phi)$ and
$\del\phi$ are the generators of an SU(2) symmetry at the boundary. The
fact
that $\cos 4\pi\phi$ can be rotated into $\del\phi$ means that the
operator is
exactly marginal, and the boundary interaction does not break the
conformal
symmetry. The resulting boundary $S$ matrix cannot depend on rapidity:
because
there are no dimensional parameters in the theory, there is no way of
making a
relativistic invariant out of a single momentum. Indeed, we see that
the
rapidity dependence in the $S$ matrix \bdrs\ goes away. This does not
mean that
the $S$ matrix is trivial since  we can scale $\theta_B$ with $\l$.
Actually, in the limit $\lambda\rightarrow 0$ this is necessary
to get a finite value of $g$, following the relation of
$g$ and $\theta_B$ (see (4.16) below).
Returning
to the
original parameter $\Theta$, we have
\eqn\bdrylimit{\eqalign{P=&{e^{\Theta/2}\over \sqrt{2\cosh\Theta}}\cr
Q=&i{e^{-\Theta/2}\over \sqrt{2\cosh\Theta}}.\cr}}
This is now a simple rotation in the $S, A$ space
\eqn\rot{\pmatrix{\cos\pi g_R&i\sin\pi g_R\cr
i\sin\pi g_R& \cos\pi g_R\cr},}
with
$$\tan \pi g_R=e^{-\Theta}.$$
This is exactly the picture obtained in \refs{\CK, \CKi,\PT}, where
$g_R$ is
the renormalized coupling constant.

\subsec{The boundary $S$ matrix for breathers}

One can easily treat similarly the boundary $S$ matrix for the $n$th
breather
\ghosi. The integrability requires that it be diagonal.
One finds amplitudes that depend on the parity of $n$:
$${1\over i}{d\over d\t}\ln R^{(2k)}(\t)=
2\sum_{l=1}^k{\cosh\t\cos(l-1/2)\pi/\lambda\over
 \cosh^2\t-\sin^2(l-1/2)\pi/\lambda}.$$
and
$${1\over i}{d\over d\t}\ln R^{(2k-1)}(\t)={1\over\cosh\t}+
2\sum_{l=1}^{k-1}{\cosh\t\cos l\pi/\lambda\over
 \cosh^2\t-\sin^2 l\pi/\lambda}.$$
The integral representation is
\eqn\breathscat{R^{(n)}(\t)
=\exp i \int_{-\infty}^{\infty} {dy\over 2y} \sin{2\l\t y\over \pi}
{\sinh ny \over \sinh y \cosh \l y}.}

\newsec{The exact boundary free energy}

Using this scattering description we can now compute the evolution of
the
boundary entropy between Neumann and Dirichlet boundary conditions.
Recall that if we consider
the theory defined by the Lagrangian \lagr\ at temperature $T$ the
free energy  reads
\eqn\parti{F=fL-T\ln (g_1g_2),}
where the bulk term $f$ is $L$ independent, $\ln g_1$ and $\ln g_2$ are
boundary entropies
associated with the two boundaries at $\sigma=0,L$. The boundary
interaction
does not affect $f$, but it causes $g_1$ to flow from its Neumann value
to
its Dirichlet value. As we will see, $g_1$ decreases, in accordance
with the
``$g$-conjecture'' of \ALii. In this section, we explicitly compute
$g_1$ as a
function
of the boundary scale by using the thermodynamic Bethe ansatz (TBA)
\refs{\YY,\Ztba}.

\subsec{The TBA equations in the presence of a boundary}

The TBA exploits the fact in an integrable model, we can find the exact
relation between the density of states $P$ and the particle density
$\rho$. (Often the quantity $P-\rho$ is called the density of holes
$\rho^h$.)
In models that are not integrable, one generally uses the free-particle
density of states and then uses perturbation theory in the interaction
parameters. However, the fact that the scattering is completely elastic
and
factorizable in an integrable model gives an exact functional relation
between
$P$ and $\rho$, known as the Bethe equation. We then write down the
free energy
in terms of $P$ and $\rho$. The particle density $\rho$ of the system
at
temperature $T$ is then found by minimizing the free energy, using the
relation
of $P$ to $\rho$.

We discuss the case when the bulk and boundary $S$ matrices are
diagonal. When
there are
$p$ different species of particle, we have the two-particle $S$ matrix
elements
$S_{rs}(\t_1-\t_2)$, where $r$ and $s$ run from $1$ to $p$. These
$S$ matrix
elements are the phase shift in the wavefunction when two particles
are
exchanged. We also have the boundary $S$ matrix elements
$R_r(\t-\t_B)$, which
gives
the phase shift when a particle of species $r$ bounces off a wall and
changes
its rapidity from $\t$ to $-\t$. We have a gas of ${\cal N}$ particles
on a
line of length $L$, with the $i$th
particle of species $r_i$.
As in the Kondo problem \PF, we map the problem onto
a line of length $2L$ ($-L<\sigma<L$)
by considering the right movers to be left movers with $\sigma < 0$.
Thus we have only
left movers scattering among themselves and off of the boundary, which
can now
be thought of as an impurity (a particle with rapidity $\t_B$). (This
trick is
common to boundary conformal field theory and can only be used in the
massless
limit.) For simplicity, we put periodic boundary conditions on the
system;
these do not change the boundary effects at $\sigma=0$.

First we write the Bethe equations. These equations quantize the set of
allowed
rapidities for
a system of left-moving particles on a circle of length $2L$ with an
impurity at $\sigma=0$.
They are obtained by collecting all the phase shifts due to
particle-particle
and particle-impurity scattering when arguments in the wave functions
are
analytically continued $\sigma\rightarrow \sigma+2L$.  Demanding that
the
wavefunction be periodic gives the constraint
\eqn\quant{e^{-2i\mu_{r_i}\exp\t_iL}\prod_{j=1,j\ne i}^{\cal N}
S_{r_ir_j}(\t_i-\t_j)R_{r_i}(\t_i-\t_B)=1 \ ,}
where $E=-P= \mu_r exp(-\theta)$ for a particle of type $r$.
Since the scattering is diagonal, we can define the densities $\rho_r$
and
$\rho_r^h$ for each species of particle, so that
$(\rho_r(\t)+\rho_r^h(\t))2Ld\t$ gives the number of allowed rapidities
between $\t$ and $\t+d\t$ for species $r$, and $\rho_r$ gives the
density of
filled states. Taking the derivative of the logarithm of \quant\ gives
one equation for every type of particle:
\eqn\sba{2\pi(\rho_r(\t)+\rho_r^h(\t))= \mu_r e^{-\t} + \sum_{s=1}^p
\varphi_{rs}\star\rho_s(\t)+ {1\over 2L}
\kappa_r(\t-\t_B),}
where $\star$ denotes convolution:
$$ f\star g(\alpha)\equiv \int_{-\infty}^{\infty} d\alpha'
f(\alpha-\alpha')
g(\alpha')$$
and
\eqn\forphi{\eqalign{\varphi_{rs}(\t)&=-i {d\over d\t}
\ln S_{rs}(\t)\cr
\kappa_{r}(\t-\t_B)&=-i {d\over d\t} \ln R_{r}(\t-\t_B).\cr}}
The effect of the boundary is seen in the last piece of \sba,
proportional to $1/L$.

We now consider the free energy
${\cal F}={\cal E}-T{\cal S}$. For the energy we have
\eqn\ener{{\cal E}=2L\int d\t\sum_{r=1}^{p}\rho_r(\t)\mu_re^{-\t}}
and entropy
\eqn\entro{{\cal S}=2L\int
d\t\sum_{r=1}^p\left[(\rho_r+\rho_r^h)\ln(\rho_r+\rho_r^h)-
\rho_r\ln\rho_r-\rho_r^h\ln\rho_r^h\right].}
(The particles act like fermions when filling levels.)
The free energy is the value of $-T\ln Z$ at the
saddle point of the partition sum or path integral: as
a function of $\rho$, ${\cal F}$ is at a miminum. Via
$${d{\cal F}\over d\rho_r}={\del{\cal F}\over \del\rho_r} +
\sum_s{\del{\cal F}\over \del\rho^h_s}{\del\rho^h_s\over
\del\rho_r}=0$$
one obtains equations for densities, using \sba\ to determine
${\del\rho^h_s/\del\rho_r}$.
These equations do not depend on the boundary term, because boundary
terms
only appear in the Bethe equations, and disappear in a variation
of the densities. Defining
$$e^{\overline\epsilon_r(\t)}\equiv
{\rho^h_r(\t)\over\rho_r(\t)},$$
one finds the well-known equations \refs{\YY,\Ztba}
\eqn\tba{\epsilon_r(\t)={\mu_r\over \mu} e^{-\t}-{1\over 2\pi}
\sum_{s=1}^p \varphi_{rs}
\star\ln(1+e^{-\epsilon_s}),}
where $\epsilon(\t)=\overline\epsilon(\t+\ln(\mu/T))$.
Boundary terms enter when expressing the free energy in terms of the
$\epsilon_r$. We find them by rewriting the energy \ener\ as
$$\eqalign{{\cal E}&=2L\int \sum_{r=1}^{p} \rho_r
\left[T\overline\epsilon_r + {T\over 2\pi} \sum_{s=1}^p \varphi_{rs}
\star\ln(1+e^{-\overline\epsilon_s})\right]\cr
&=2LT\int \left[\sum_{r=1}^{p}\rho_r\overline\epsilon_r
+\sum_{s=1}^p \left(\rho_s+\rho_s^h - {\mu_s\over 2\pi}e^{-\t}
-{1\over 4\pi L}\kappa_s \right)
\ln(1+e^{-\overline\epsilon_s})\right]\cr
&=T{\cal S} - \int \sum_{r=1}^{p}\left[{TL\mu_r\over \pi} e^{-\t} +
{T\over 2\pi}\kappa_r\right]\ln(1+e^{-\overline\epsilon_r}),\cr}$$
where we used \tba\ to get to the first line, \sba\ to get to the second, and
\entro\ to get to the third.
Thus we have
\eqn\freeen{{\cal F}={\cal F}_{bulk}-
T\int {d\t\over 2\pi}\sum_{r=1}^{p}
\kappa_{r}(\t-\ln(T/T_B))\ln(1+e^{-\epsilon_r(\t)}).}
It is a general result that ${\cal F}_{bulk}=
-{\pi c\over 6}T^2L$ in a massless bulk theory,
where $c$ is the central charge of the
conformal field theory \BCNA.

Although the equations \tba\ for $\epsilon(\t)$ cannot
be solved explicitly
for all temperatures, the free energy is
easy to evaluate as $T\to 0$ and $T\to\infty$,
as we
will show in sect.\ 4.3. Moreover, one can extract the
analytic
values of critical exponents by looking at the form of the expansions
around
these fixed points. Also, they are straightforward to solve numerically
for any
$T$.

Several notes of caution
are necessary. At the order we are working, the formula \entro\ for the entropy
is not quite
correct, because
there are $1/L$ corrections to the Stirling formula used in its
derivation.
Also,
at this order, the logarithm of the partition function is not ${\cal
E}-T{\cal
S}$: it depends not only on the saddle point
value of the sum over all states, but also on fluctuations. Their net
effect is that we cannot compute the $g$ factors from ${\cal F}$ alone.
However, both of these corrections are subleading contributions to the
bulk
free energy, and do not depend on the boundary conditions. Therefore   we can
still compute {\bf differences} of $g$
factors from
${\cal F}$; the corrections are independent of the boundary scale
$\t_B$ and
cancel out of the difference \FSiii.

\subsec{The exact boundary free energy for $\l$ integer}

In this section we consider only the case $\lambda$ a positive integer.
At these values, a major simplification takes place: the bulk
scattering \sll\
is diagonal. The boundary scattering still is not, but we can redefine
our
states to be $|\pm\>\equiv (|S\>\pm |A\>)/\sqrt{2}$ as in the previous
section,
so that the boundary scattering is now diagonal:
\foot{If $\lambda$ is even, this actually makes the bulk scattering
completely
off-diagonal (e.g.\ $|++\>$ scatters to $|--\>$), but the TBA equations
turn
out the same.}
$$\eqalign{R^+(\t)=P+Q&=\cosh\left({\l\t\over 2}
-{i\pi\over 4} \right) e^{i\pi/4} R(\t)\cr
R^-(\t)=P-Q&=\cosh\left({\l\t\over 2}
+{i\pi\over 4}\right) e^{-i\pi/4} R(\t)\cr}$$

Using the analysis of the previous subsection, we can read off the
answer.
There are $\l +1$ particles: the breathers, which we index $n=1\dots
\l-1$, and
the $\pm$ particles. We define the Fourier transform as
$$\tilde f(y)=
\int^{\infty}_{-\infty} {d\t\over 2\pi} e^{i 2 \l\t y/\pi }
f(\t).$$
The bulk kernels $\phi$ are well known \KM; they can be written in the
form
\eqn\keri{\eqalign{&\tilde \varphi_{nl}=\delta_{nl}
-2{\cosh y \cosh(\l-n)y \sinh ly
\over \cosh \l y \sinh y}
\qquad n,l=1\dots \l-1;\ n\ge l\cr
&\tilde \varphi_{n,\pm}=-{\cosh y
\sinh  ny \over \cosh \l y
\sinh y}\cr
&\tilde \varphi_{\pm,\pm}=\tilde \varphi_{\pm,\mp}
=-{\sinh  (\l-1)y \over
2 \cosh \l y \sinh y},\cr}}
with $\varphi_{rs}=\varphi_{sr}$. The boundary kernels follow from
\forphi, and are
\eqn\kerii{\eqalign{\tilde \kappa_n&=
{\sinh n y \over 2 \sinh y \cosh \l y}\cr
\tilde \kappa_-&= {\sinh(\l-1)y \over 2 \sinh 2y\cosh \l y }+
{1\over 2 \cosh y} \cr
\tilde \kappa_+&= {\sinh(\l-1)y \over 2 \sinh 2y\cosh \l y }.\cr}}
The equations for the $\epsilon_r$ can be written in a much simpler
form by
using a few trigonometric indentities. One finds
\eqn\tbaii{\epsilon_r= K\star\sum_s N_{rs}\ln(1+e^{\epsilon_s}),}
where the label $s$ runs over breathers and $\pm$, the kernel
$K(\t)={1/2\pi\cosh\t}$, and $N_{rs}$ is the
incidence matrix of the following diagram

\bigskip
\noindent
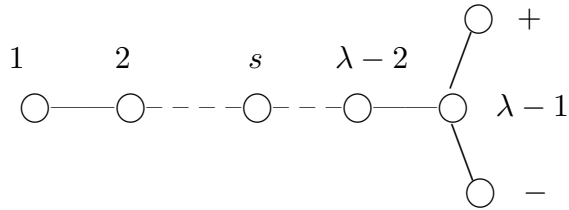
\centerline{\hbox{\rlap{\raise28pt\hbox{$\hskip5.5cm\bigcirc\hskip.25cm
+$}}
\rlap{\lower27pt\hbox{$\hskip5.4cm\bigcirc\hskip.3cm -$}}
\rlap{\raise15pt\hbox{$\hskip5.1cm\Big/$}}
\rlap{\lower14pt\hbox{$\hskip5.0cm\Big\backslash$}}
\rlap{\raise15pt\hbox{$1\hskip1cm 2\hskip1.3cm s\hskip.8cm \l-2$}}
$\bigcirc$------$\bigcirc$-- -- --
--$\bigcirc$-- -- --$\bigcirc$------$\bigcirc$\hskip.3cm $\l-1$ }}

\bigskip
\noindent
The dependences on the mass ratios seems to have disappeared from \tbaii,
but they
appear as an asymptotic condition: the original equations \tba\
indicates that
the solution must satisfy
$$\epsilon_r \to {\mu_r\over \mu} e^{-\t}
\qquad \hbox{as}\ \t\to -\infty.$$

To recapitulate: the impurity contribution to the free energy is
\eqn\fimp{{\cal F}_{imp}= -T\int {d\t\over 2\pi}\sum_r
\kappa_{r}(\t-\ln(T/T_B))\ln(1+e^{-\epsilon_r(\t)})), }
where the $\kappa_r$ are given by \kerii, and the $\epsilon_r$ are the
solutions of the non-linear integral equations \tbaii.

At non-integer values of $\lambda$, the bulk $S$ matrix is not
diagonal.
As a result both bulk and boundary scattering have to be
simultaneously diagonalized. It is known how to do this in the bulk
\TS, and
the inclusion of the boundary can be presumably be acomplished
by using the quantum inverse scattering method along the lines of \SK\
and
\FSiii. The simplest values to do should be $\l=1/t$ with $t$ integer,
where
there are no breathers, and the bulk TBA system is given by \tbaii\
with
different conditions for $\t\to -\infty$. In fact, using the ideas
of \FSiii\
and the ``duality'' of TBA equations for $\lambda=t$ and
$\lambda={1\over t}$ we have been able to conjecture a TBA
for the latter case that reproduces all expected results
for our flow. For simplicity we do not present this TBA here.

\subsec{The Neumann and Dirichlet limits}

We can evaluate the impurity free energy explicitly in several limits.
In the IR limit $T/T_B\rightarrow 0$ the integral is dominated by
$\t\rightarrow-\infty$
where the source terms in \tba\ become very big. Hence
$\epsilon_r(-\infty)=\infty$
and the impurity free energy vanishes in this limit.
In the UV limit $T/T_B\rightarrow \infty$ the integrals
are dominated by the region where $\t$ is large
so that the source terms disappear
in \tba\ and the $\epsilon_r$ go to constants:
\eqn\epslim{x_n\equiv e^{\epsilon_n(\infty)}=(n+1)^2-1 ;\qquad \quad
x_\pm=\l.}
Therefore we obtain
\eqn\deltag{\eqalign{\ln{g_N\over g_D}=&\left.
{-{\cal F}_{imp}\over T}\right|_{UV}-\left.{-{\cal F}_{imp}
\over T}\right|_{IR}\cr
=&\sum_{n=1}^{t-2}
I^{(n)}\ln(1+1/x_n)+(I^{(+)}+I^{(-)})\ln(1+1/x_\pm),\cr}}
where
$$I^{(r)}\equiv\int {d\t\over 2\pi}\kappa_r(\t)=\tilde\kappa(0).$$
Thus $I^{(n)}=n/2$ and $I^{(+)}+I^{(-)}=\l/2$, and we find
\eqn\deltagi{\eqalign{\ln{g_N\over g_D}&={\l\over 2}\ln{\l+1\over
\l}+\sum_{n=1}^{\l-1}{n\over 4}\ln{(n+1)^2\over n(n+2)}\cr
&={1\over 2}\ln (\l+1).\cr}}
This is agreement with the ratio calculated from conformal field
theory, as
detailed in the Appendix.

We can also find the dimension of the perturbing operators.
{}From the equations \tba\ one deduces \Zamo\ the following expansions
for
$T/T_B$ large:
$$Y_r(\t)=e^{\epsilon_r(\t)}=\sum_j Y_r^{(j)}e^{-2j\l\t/(\l+1)}.$$
As a result it is straightforward to see that near $g$=0$,
{\cal F}$ can be expanded in powers
of $(T_B/T)^{2\l/(\l+1)}$.
On the other hand we expect ${\cal F}$ to be an analytic
function of $g^2$. Hence
\eqn\coupl{g\propto \left(\mu e^{-\t_B}\right)^{\l/(\l+1)}.}
This agrees with the conformal result that the perturbing
operator $\cos[ \beta\phi(0)/2]$ has boundary dimension
$x=1/(\l+1)=\beta^2/8\pi$.
In the IR limit of $T/T_B$ small, one can expand out the kernels
$\kappa_r$ in
powers of $\exp(\t_b-\t)$. This leads to the fact
that the irrelevant operator which
perturbs the Dirichlet boundary conditions has dimension $x$=2. This is
the
energy-momentum tensor. (We note that there is another irrelevant
operator in
the spectrum with dimension $x=\l+1$, which for $0\le \l<1$ is the
appropriate
perturbing operator \KF.)

\newsec{Future directions}

We see that at $\l=0$, the boundary
$S$ matrix is a constant for all $\theta$. This is
expected,
since we can continuously interpolate from Neumann to Dirichlet without
destroying the conformal symmetry. Using the conformal symmetry, one in
fact is
able to calculate the full partition function along this interpolation
\refs{\CK-\PT}, not just at the Neumann and Dirichlet points as in the
Appendix. It would be interesting to precisely understand how to map
the
scattering basis discussed
in \CKi\ to the scattering basis used here. This
would
provide a great deal of insight into the long-confusing relation
between the
states arising from the Bethe Ansatz to those in the Virasoro towers of
a
conformal field theory (see \Melzer, for example).

The quantum $S$ matrix for Dirichlet  boundary conditions
$\phi(0)=\phi_0$ can be
derived explicitly by applying the Bethe ansatz to an underlying
lattice model,
the XXZ spin chain with a boundary magnetic field \FSiii.
Presumably the more general $S$ matrix studied here
corresponds to the XXZ chain with  boundary terms of the
type $\sigma^{\pm}$ that break the $U(1)$ symmetry
\ref\DV{H.J. de Vega and A. Gonzalez Ruiz, ``Boundary K matrices
for the XYZ, XXZ and XXX spin chains'', LPTHE-93-29,
hep-th/9306089.}.

By expanding the partition function of our system in powers of $g$
we obtain a one-dimensional neutral Coulomb gas of particles on a
circle, whose solution is therefore indirectly provided by the TBA.
Previously, such a Coulomb gas had
been solved for only one type of charge.

Perhaps the most interesting future direction is the application of
these
results to the experimentally realizable system of tunnelling between
fractional
quantum-Hall edges \QHE. We will report on this application soon
\FLS.

\bigskip

\leftline{\bf Acknowledgements:}

We thank the many participants at the conference SMQFT 94 for discussions.
This work was supported by the Packard
Foundation, the
National Young Investigator program (NSF-PHY-9357207) and  the DOE
(DE-FG03-84ER40168).

\appendix{A}{Neumann and Dirichlet partition functions}
We here use standard conformal theory techniques to calculate the
partition function of our quantum system on a cylinder of length $L$
and circumference $1/T$ and Neumann or Dirichlet boundary conditions
at the ends.

With Dirichlet boundary conditions on both sides, winding modes are not
allowed
around the cylinder, but they are allowed along it. Thus the field
$\phi$
obeys $\phi(\sigma=0)=\phi(\sigma=L)+n 4\pi/\beta $
with $n$ an integer. (In the usual conventions,
this corresponds to a bosonic radius of $\sqrt{4\pi}/\beta$.)
Hence \EA
\eqn\znn{Z^{DD}={1\over\eta(w)}\sum w^{8\pi n^2/\beta^2},}
where $w=e^{-\pi /LT}$. After modular transformation one finds
\eqn\znni{Z^{DD}=\left({\beta^2\over 16\pi}\right)^{1/2}
{1\over\eta(q^2)}\sum q^{\beta^2n^2/16\pi},}
where $q=e^{-2\pi LT}$.
With Neumann on one side and Dirichlet on the other, no winding modes
of any
sort are allowed. The energy, however, is quantized with half-integer
eigenvalues. Using the Jacobi triple product formula gives
\eqn\znd{Z^{ND}={1\over \eta(w)}\sum w^{(n-1/2)^2/4},}
which after modular transformation is
\eqn\zndi{Z^{ND}={1\over \sqrt{2}\eta(q^2)}\sum (-1)^n q^{2n^2}.}
By taking the ratio in the limit of large $L$, we find
\eqn\gneu{{g_D\over g_N}=\left(\beta^2\over 8\pi\right)^{1/2}.}
in agreement with \deltagi.

\listrefs
\bye